\documentclass[grl]{agutex}

\usepackage{graphicx}
\authorrunninghead{MILLER ET AL.}

\titlerunninghead{PMCS IMAGING}

\authoraddr{Corresponding author: A. D. Miller,
Columbia University
550 W. 120th St. Mail Code 5247
New York, NY 1002 USA.
(amiller@columbia.edu)}

\begin{document}


\title{Stratospheric Imaging of Polar Mesospheric Clouds: A New Window on Small-Scale Atmospheric Dynamics}




\authors{
A. D. Miller,\altaffilmark{1}
D. C. Fritts,\altaffilmark{2}
D. Chapman,\altaffilmark{1}
G. Jones,\altaffilmark{1}
M. Limon,\altaffilmark{1}
D. Araujo,\altaffilmark{1}
J. Didier,\altaffilmark{1}
S. Hillbrand,\altaffilmark{3}
C. B. Kjellstrand,\altaffilmark{1}
A. Korotkov,\altaffilmark{4}
G. Tucker,\altaffilmark{4}
Y. Vinokurov,\altaffilmark{4}
K. Wan,\altaffilmark{2} and
L. Wang\altaffilmark{2}
}

\altaffiltext{1}{Columbia University, New York, NY 10027.}

\altaffiltext{2}{GATS Inc., Boulder, CO 80301.}

\altaffiltext{3}{California State University, Sacramento, CA 95819.}

\altaffiltext{4}{Brown University, Providence, RI 02912.}



\begin{abstract}
Instabilities and turbulence extending to the smallest dynamical scales play important roles in the deposition of energy and momentum by gravity waves throughout the atmosphere. However, these dynamics and their effects have been impossible to quantify to date due to lack of observational guidance. Serendipitous optical images of polar mesospheric clouds at $\sim$82 km obtained by star cameras aboard a cosmology experiment deployed on a stratospheric balloon provide a new observational tool, revealing instability and turbulence structures extending to spatial scales $<$ 20~m. At 82 km, this resolution provides sensitivity extending to the smallest turbulence scale not strongly influenced by viscosity: the "inner scale" of turbulence, $l_0\sim$10($\nu^3$/$\epsilon$)$^{1/4}$. Such images represent a new window into small-scale dynamics that occur throughout the atmosphere but are impossible to observe in such detail at any other altitude. We present a sample of images revealing a range of dynamics features, and employ numerical simulations that resolve these dynamics to guide our interpretation of several observed events.
\end{abstract}

%
%

%

\begin{article}

\section{Introduction}

Earth's atmosphere hosts a wide range of motions, from mean flows, planetary waves, and tides on global scales, to instabilities and turbulence on scales of meters or less \citep{Andrews1987, Wyngaard2010}. At intermediate scales, $\sim$10’s to 1,000’s of km, atmospheric gravity waves (GWs) play key roles in defining Earth's weather and climate due to their many effects, especially their efficient vertical transport of horizontal momentum at smaller spatial scales \citep{Bretherton1969, Holton1982, GarciaSolomon1985, Haynes1991}. GW momentum deposition requires dissipation, and the dominant processes below $\sim$100 km are small-scale instabilities and turbulence \citep{FrittsAlexander2003}. Importantly, these small-scale dynamics influence the spatial scales and variability of GW dissipation and momentum deposition throughout the atmosphere \citep{Fritts2013}. They also play central, but poorly understood, roles in defining the spectrum of GWs penetrating to even higher altitudes \citep{Vadas2007}.  

Global weather and climate models typically do not resolve the small-scale GWs that contribute the majority of momentum transport, nor the dynamics that drive their dissipation \citep{FrittsAlexander2003}. However, these effects must be included via parameterization in order to describe global GW influences \citep{McFarlane1987, WarnerMcIntyre1996, Hines1997}.
Because these small-scale dynamics are poorly understood, their current descriptions in weather and climate models are qualitative, simplistic, and acknowledged to have deficiencies that limit model and forecast accuracy \citep{Kim2003, McLandress2012, McIntyre1990}. 

Our poor understanding of GW instability dynamics and momentum deposition is due to their complexity and diversity. These dynamics are strongly nonlinear and are further complicated because they often occur in multi-scale flows involving superpositions of motions having various scales and frequencies \citep{LombardRiley1996, SonmorKlaassen1997, Andreassen1998, Fritts2009a, Fritts2009b, Fritts2013}. GW dynamics also play similar roles in the structure and variability of other stratified fluids, including oceans, lakes, other planetary atmospheres, and stellar interiors \citep{Thorpe2005}.

Perhaps the best region of the atmosphere in which to study the instability and turbulence dynamics accounting for GW momentum deposition is near the mesopause. This is because the mesopause region ($\sim$80--100 km) experiences frequent large GW amplitudes and strong responses to momentum deposition, hence the GW responses are often easier to quantify than at lower altitudes. As a result, many measurement techniques have been employed to probe these dynamics at altitudes ranging from $\sim$70--110 km, including airglow imagers, radars, lidars, in-situ instruments, and others \citep{TaylorHapgood1988, Taylor1995, Lubken1997, Yamada2001, Hecht2004, Rapp2004, Strelnikov2006, Williams2006, Lehmacher2007, Pfrommer2009, Hecht2014}. 

A unique, additional benefit for studies near the mesopause is that GW dynamics contribute to formation of a thin layer of ice clouds known as polar mesospheric clouds (PMCs) at $\sim$80--85 km altitudes \citep{Thomas1991, Hervig2001, Russell2009, Gordley2009} that act as very sensitive tracers of the various GW, instability, and turbulence dynamics \citep{Witt1961, Baumgarten2014}. A primary reason PMCs form is because GW momentum deposition drives an induced mean inter-hemispheric circulation from the summer to the winter mesopause that results in mean upwelling that is maximum at summer polar latitudes \citep{Holton1982, GarciaSolomon1985}. This upwelling induces strong cooling and causes the polar summer mesospause to be the coldest region on Earth, e.g., mean temperatures of $\sim$130 K at $\sim$88 km \citep{vonZahnMeyer1989, Lubken1999}. PMCs tend to be much thinner than the airglow layers, and are thus more sensitive to small features. For these reasons, PMC images, obtained from the ground for over 60 years and more recently from space, have provided useful clues to the important GW and instability dynamics near the mesopause \citep{Witt1961, Fritts1993, Jensen1994, Russell2009, Rusch2009, Baumgarten2014}. 

PMC imaging from the ground and space is constrained by resolution and viewing geometry.  Ground-based viewing is limited to shallow viewing angles (resulting in a viewing range from the ground site of $\sim$300--600 km) due to sky background and weak PMC brightness. This results in image distortion and blurring by atmospheric turbulence \citep{Witt1961, Baumgarten2014}. Similarly, wide satellite imager field of views (FOVs), large range ($\sim$600--1000 km), and finite integration times yield minimum observable scales of $\sim$10 km. Thus, new PMC imaging from a stratospheric balloon exhibiting apparent GW instability and turbulence structures at spatial scales as small as $\sim$10--20 m is a major advance. 

Here we explore the confluence between the new high-resolution imaging of PMCs and high-resolution numerical simulations of representative instability and turbulence dynamics accompanying idealized GW breaking and multi-scale flows. Our goals are 1) to demonstrate the potential for PMC images to reveal key small-scale dynamics that provide evidence of GW dissipation and momentum deposition and 2) to use numerical simulations to provide interpretations of some of these dynamical features and explore their implications for GW forcing of the mesopause region.  The observations and modeling results employed for this study are described briefly below. We then discuss their applications in the interpretations of PMC images suggesting idealized and more complex dynamics. 

\section{Observations}

The E and B Experiment (EBEX) was designed to measure polarization in the cosmic microwave background \citep{ebex}. EBEX flew on a balloon at approximately 35 km over Antarctica between December 29, 2012 and January 9, 2013, coincident with the beginning of the summer hemisphere PMC season. Two star cameras employed for precise pointing measurements obtained over 40,000 images \citep{Chapman2014,Chapman2015}, approximately half of which show significant PMC activity. Each camera had $1536 \times 1024$ pixel resolution, and a FOV of $4.1^{\circ}\times2.7^{\circ}$. The cameras typically viewed at zenith angles of 36 degrees, implying a range to the PMC layer of $\sim$62~km, $\sim$5--10 times closer than is possible with ground-based imaging \citep{Baumgarten2014}.  

The star camera images exhibit a wide range of dynamics, from larger-scale (several km) coherent structures to features as small as the star camera pixel resolution (10--20 m). Notably, the smallest observed scales are comparable to the smallest scales within the inertial range of turbulence at the PMC altitude \citep{Lubken1997, Lubken2002, Rapp2004}. The combination of resolution spanning a wide range of scales and 2D viewing of flow features more nearly overhead (and without significant distortion) provides a unique window on instability and turbulence dynamics that is not available at any other altitude. 

\section{Numerical simulations}     

Continuing advances in computational capabilities now allow numerical modeling of GW, instability, and turbulence dynamics for idealized and multi-scale flows having similar character to those observed at mesopause altitudes. In particular, these include resolution of the turbulence inertial range for realistic GW spatial scales, kinematic viscosity, $\nu$, and Reynolds numbers, Re=UL/$\nu$, where U and L are characteristic velocity and length scales of the GW field \citep{Fritts2009a, Fritts2009b, Fritts2013}. 

The model describing GW dynamics and their implied PMC structures solves the nonlinear Navier-Stokes equations in three dimensions employing the methods and initial conditions described in previous papers \citep{Fritts2009a, Fritts2013}. Because of the range of dynamics revealed by the PMC images, simulations of both idealized GW breaking and a multi-scale flow (e.g., a superposition of various motions having different character) are employed.  

The idealized GW breaking case \citep{Fritts2009a} employs an initial GW amplitude 
$a=u$'$/(c-U)=1.1$ (just above the overturning amplitude $a=1$), intrinsic frequency $\omega=0.32N$, vertical wavelength $\lambda_z=10$~km, and PMC full-width half-maxima of 1.5 and 0.5~km, where $N$, $c$, $u$', and $U$ are the buoyancy frequency, GW phase speed, horizontal perturbation velocity, and mean wind speed, respectively. Small initial perturbations trigger instabilities that amplify with time and lead to intensified shear layers. These roll up into sequences of vortex rings and trailing vortices that drive the subsequent transition to turbulence. Embedded PMC layers reveal various aspects of these dynamics, depending on the PMC locations and times within the GW breaking event. 

The multi-scale GW case \citep{Fritts2013} employs an initial GW with $a=0.5$ and a superposed oscillatory mean shear having a vertical wavelength $\sim$5 times smaller than the GW. These fields interact strongly, resulting in a superposition of GWs having evolving amplitudes and phase structures. This superposition yields multiple sites of instabilities, including  local GW breaking, Kelvin-Helmholtz shear instabilities (KHI), and fluid intrusions, spanning many buoyancy periods.

The simulations described above include a wide range of GW and instability dynamics, scales, and intensities throughout their evolution. Each also describes advection of a PMC layer having an undisturbed initial Gaussian brightness distribution in altitude with variable depths and initial altitudes in order to explore the PMC responses to the various events and identify those that compare favorably with the PMC images.

\section{Results and Discussion}

Figure 1 shows two EBEX PMC images suggesting an initial GW breaking front with trailing vortices and successive coherent vortex rings (panels A and C). Panels B and D show modeled PMCs from the idealized GW breaking simulation at corresponding instability stages. The bright regions in each correspond to regions in which PMC brightness has accumulated along the line of sight (LOS) relative to where it has thinned by being stretched and viewed more in a normal direction. Accumulation typically occurs where the GW phase structure is overturning or where instabilities concentrate vorticity (and PMC brightness where they occur together) accompanying the dynamics that drive the transition to full turbulence. Both the observed and modeled GW breaking fronts exhibit regions of high PMC brightness where the PMC layer is overturning. Both fronts also exhibit undulations and trailing vortices behind the fronts (panels A and B, bottom) accompanying initial frontal instability dynamics, and an absence of such features ahead of the front (panels A and B, top right). In each case, there is little evidence of pre-existing turbulence (see discussion below), though the right portion of the observed PMC exhibits a more advanced transition to turbulence. The two images of vortex rings (panels C and D) suggest darker rings against a brighter background, as if the rings are penetrating into a region of greater PMC brightness from above \citep{Fritts2009b}. In both cases, vortex rings appear in succession. The modeled PMC also reveals that the rows of vortex rings occur side-by-side, but the EBEX PMC image has a FOV that is too limited to reveal adjacent rows. Based on the EBEX image scales, vortex ring diameters are $\sim$2~km and suggest a GW $\lambda_z\sim$5--10~km, based on the model results. Such a GW would necessarily have a large amplitude and contribute significant momentum deposition accompanying breaking and dissipation. The observed vortex rings also appear to occur in a region of pre-existing turbulence, given the less coherent features seen at smaller scales throughout this image. 

Figure 2 shows EBEX PMC images of more turbulent fronts (panels A and C-E). The sequence of EBEX PMC images in panels C-E are spaced at $\sim$30-s intervals and illustrate an ability to track the evolution of specific features where successive common viewing geometries occur. Such turbulent fronts do not occur in simulations of idealized GW breaking performed to date. However, they are very similar to features seen in the multi-scale simulation discussed above \citep{Fritts2013}. Modeled PMC displays are shown for comparison in panels B and F. Key features common to the observed and simulated PMCs include the following: 1) front penetration into apparently more quiescent air (towards the lower right in A and B and towards the upper right in C-F), 2) apparent transitional dynamics exhibiting coherent, larger-scale features trailing the fronts and elongated nearly normal to the fronts, and 3) less coherent turbulence structures further behind the fronts. The EBEX PMC images reveal spatial scales ranging from $\sim$1 km or greater to as small as $\sim$20 m. Thus, some portions of these fields are strongly turbulent at the times of these images. Visible features at $\sim$20-m scales suggest a comparable turbulence inner scale, $l_0\sim$10($\nu^3$/$\epsilon$)$^{1/4}$, and an implied energy dissipation rate of $\epsilon$ $\sim 0.05$~${\rm m}^2{\rm s}^{-3}$, which is at the large end of the range of values measured by in-situ instruments at these altitudes \citep{Lubken2002, Rapp2004}.  

EBEX PMC images in Figure 3 exhibit additional features that are suggestive of other dynamics observed to occur in the multi-scale simulation described above. The EBEX image in panel A reveals an apparent bright turbulent wake from several localized sources at small scales that is followed by very dark regions (likely from outside the initial PMC layer) that are at an earlier stage of turbulence development. The modeled PMC image (panel B) exhibits very similar features and accompanies an active region of GW breaking within the multi-scale flow that spans a significant depth. The EBEX image in panel C reveals multiple apparent cusp-like features that closely resemble those seen in the multi-scale model PMC field at the lower right. Features in the modeled image occur at the upper edge of a significant region of GW breaking at small spatial scales. The sharper and smaller-scale features in the EBEX image (as small as $\sim$20 m) suggest that this breaking event was also at a higher Reynolds number (and larger implied GW scale) than that modeled, again indicating local energy dissipation rates comparable to those inferred above. The implied Reynolds number is perhaps $\sim$10 times or more larger, and the corresponding GW $\lambda_z$ would be $\sim$10--20~km.    

Figures 4 and 5 show eight EBEX PMC images exhibiting intriguing features for which we have not yet found comparable structures in our numerical results. Figure 4A and 4C reveal laminar features that appear to be extended and intertwined vortices. The vortices may accompany KHI occurring at very thin shear layers and having horizontal wavelengths $\lambda_h$ $\sim$30--500~m. Similar features occur in KHI seen in PMCs from the ground, but only at significantly larger $\lambda_h$ \citep{Baumgarten2014}. These features appear to occur in environments where the background turbulence is very weak, because of the occurrence of apparent KHI at very small scales. In contrast, Figures 4B and 4D indicate strongly turbulent flows that nevertheless exhibit extended coherent features at larger and smaller scales. As above, the turbulent features extending to very small scales at right imply quite large energy dissipation rates in these events. 

Figure 5 shows an array of dynamics containing both laminar and turbulent features. Note in panel A the transition between the laminar features at upper left and the turbulence at lower right. Panel B shows what appear to be features extending roughly along both diagonals, indicating the possibility that we are viewing two layers simultaneously. Close inspection suggests a potential for some interactions between the two layers, implying that they were in close proximity. Panels C and D exhibit both smaller-scale turbulence and larger-scale wave-like features at horizontal scales of $\sim$1--3~km that could be either small-scale GWs or KHI modulating the turbulent background.  

\section{Summary}
 
High-resolution PMC images obtained from a stratospheric balloon platform provide evidence of small-scale GW instability and turbulence dynamics that play central roles in the deposition of energy and momentum throughout the atmosphere. With the aid of numerical modeling, we are able to 1) identify features indicative of specific instability dynamics, 2) recognize turbulence features extending to the smallest turbulence scales, 3) approximate associated turbulence intensities, and 4) estimate likely GW scales, amplitudes, and momentum fluxes based on observed instability types and scales. We expect further analyses of PMC images, and identification of specific dynamics using numerical modeling, will significantly advance our understanding of the character and diversity of multi-scale dynamics at PMC altitudes of relevance throughout the atmosphere. A better understanding of these dynamics will enable a more quantitative characterization of GW momentum deposition at small scales and potentially contribute to improved parameterizations of these dynamics in weather and climate models. 

Further analyses of these PMC images will include tracking the morphologies of small-scale features in time and comparisons of these images with coincident images taken by the Cloud Imaging and Particle Size (CIPS) instrument aboard the Aeronomy of Ice in the Mesosphere (AIM) satellite to identify their larger-scale context. This work has also led to the design of a dedicated experiment that expands the technique to obtain continuous imaging of PMC features with similar resolution, but larger FOVs enabling quantification of entire GW and instability events. 


%
%
%
%
%

\begin{acknowledgments}
This work was supported by NASA grants NNX08AG40G and NNX07AP36H for the development, deployment, and observations of EBEX; and NSF grants AGS-1242949 and AGS-1261619 for the modeling support. We thank Columbia Scientific Balloon Facility for their support of EBEX, and the BLAST team for sharing their attitude control system and knowledge. A.D.M. acknowledges internal support from Columbia University. J.D. acknowledges support from a NASA NESSF NNX11AL15H. We thank Gary Thomas, Aimee Merkel, Marvin Geller, and David Rind  for useful conversations. Kyle Helson, Ant\'{o}n Lizancos, Thuy Vy Thi Luu, Theodore Macioce, and Charles Zivancev performed analyses that helped to frame the scope of this project. The images used in this paper are available at http://cosmology.phys.columbia.edu/pmc/
\end{acknowledgments}

\end{article}
%
%
%
%
%
 
\begin{figure}[t]
\centering
\noindent\includegraphics[width=1.0\textwidth, angle=0]{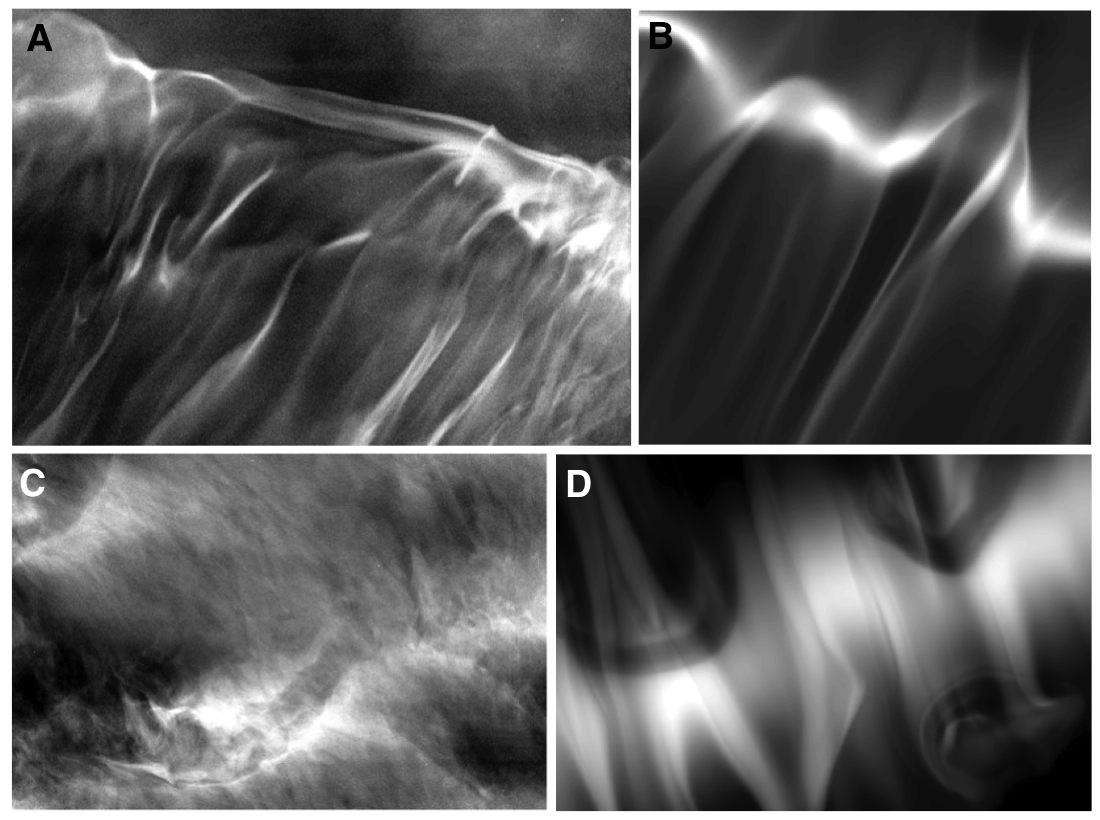}
\caption{Comparisons of EBEX PMC images and idealized GW breaking PMC simulations. Panel A shows an EBEX image (4.1 km x 3.3 km) of what appears to be a GW breaking front. A feature with the same morphology appears in the modeled breaking GW shown in panel B. GW propagation in each case is towards the upper right.  Panel C shows an EBEX PMC image (4.1 km x 3.3 km) exhibiting semi-circular features suggestive of successive laminar vortex rings in background turbulence. The corresponding result from the GW breaking simulation (panel D) shows the response to true vortex rings accompanying GW breaking in a laminar background \citep{Fritts2009b}.}
\label{fig:gravity_wave_breaking}
\end{figure}
 
\begin{figure}[t]
\centering
\noindent\includegraphics[width=1.0\textwidth, angle=0]{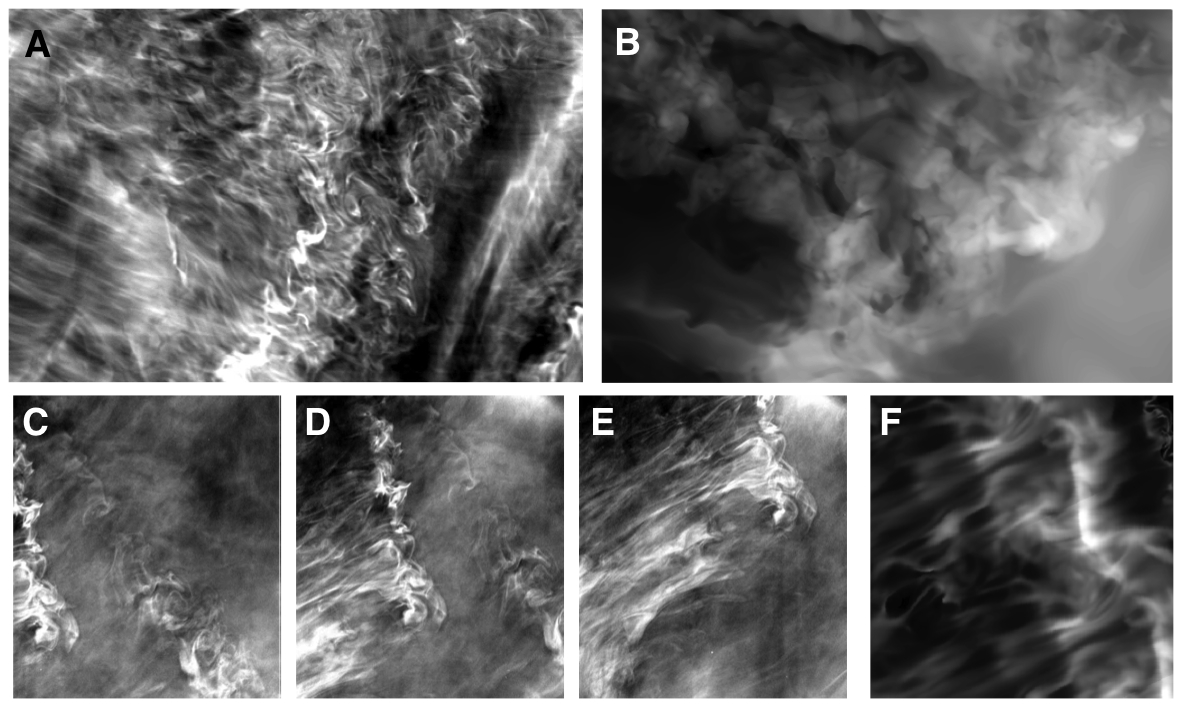}
\caption{Comparisons of EBEX PMC images (panels A and C-E) and  multi-scale simulation PMCs (B and F) showing turbulent intrusion events. Panel A shows a single intrusion event, while panels C-E show three successive views (3.3 x 3.1 km FOV) of a second event separated by $\sim$30-s intervals. Panels B and F show two simulated intrusions penetrating into quiescent air. The intrusion directions of motion are towards the lower and upper right in the upper and lower images, respectively.}
\label{fig:vortex_rings}
\end{figure}

\begin{figure}[t]
\centering
\noindent\includegraphics[width=1.0\textwidth, angle=0]{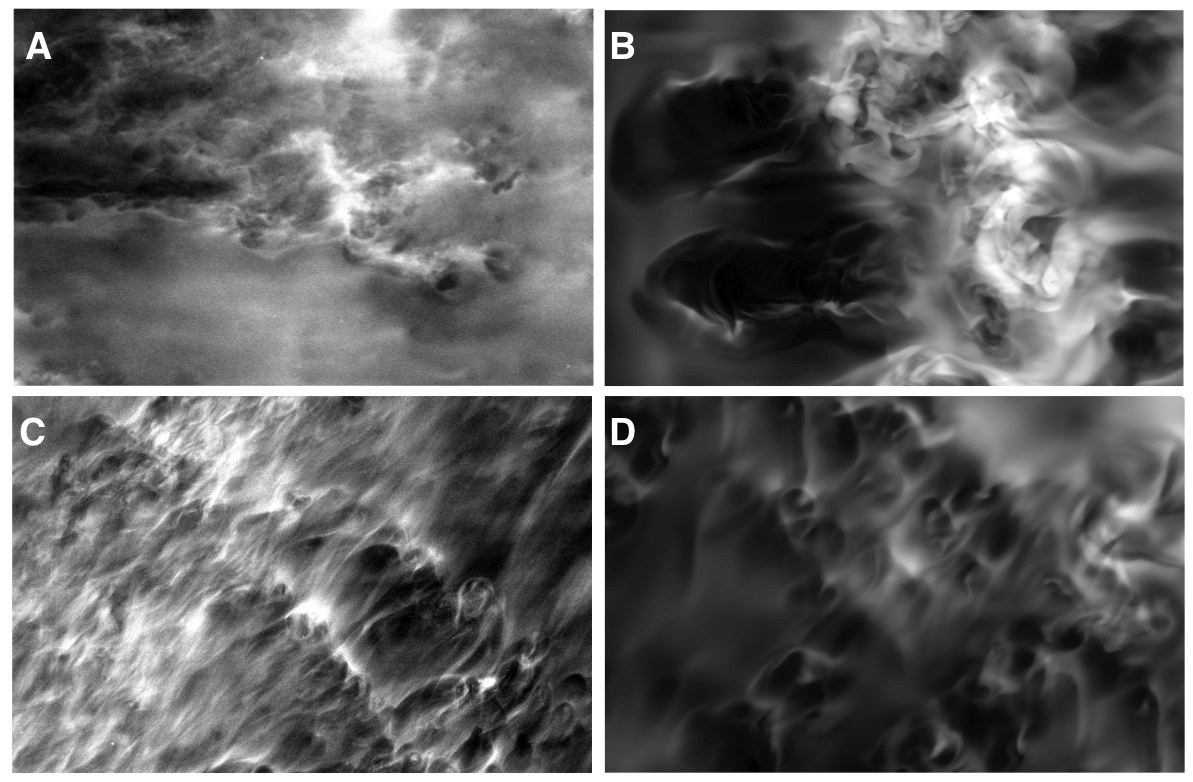}
\caption{Comparisons of EBEX PMC images (4.1 x 3.3 km FOV) and those from the multi-scale simulation showing turbulence transitions. Panels A and B show apparent turbulent wakes from localized source regions seen by EBEX (A) and in the simulation (B). Panels C and D show very similar cusp-like features seen by EBEX (C) and in the simulation (D). The multi-scale simulation indicates that the cusp-like structures occur at the upper edge of an extended region of small-scale vortex structures following their transition to turbulence.}
\label{fig:images_vs_sims}
\end{figure}

\begin{figure}[t]
\centering
\noindent\includegraphics[width=1.0\textwidth, angle=0]{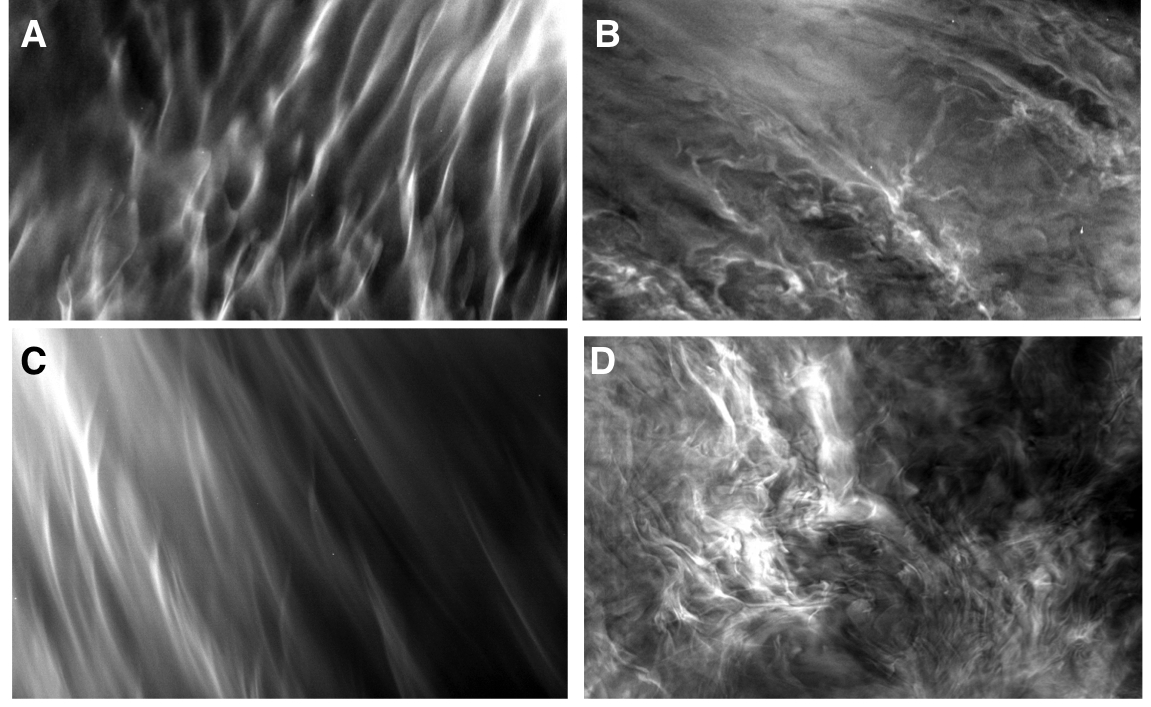}
\caption{Four EBEX images, top row a 4.1 x 3.0 km and bottom row 4.1 x 3.3 km FOV, illustrating a range of intriguing dynamics of currently unknown origin. These range from laminar features exhibiting vortex advection and intertwining (panels A and C) to highly structured turbulent regions having strong spatial variability (panels B and D). Note the extremely fine-scale features (with widths as small as $\sim$15 m) evident in the turbulent images.}
\label{fig:other_dynamics}
\end{figure}

\begin{figure}[t]
\centering
\noindent\includegraphics[width=1.0\textwidth, angle=0]{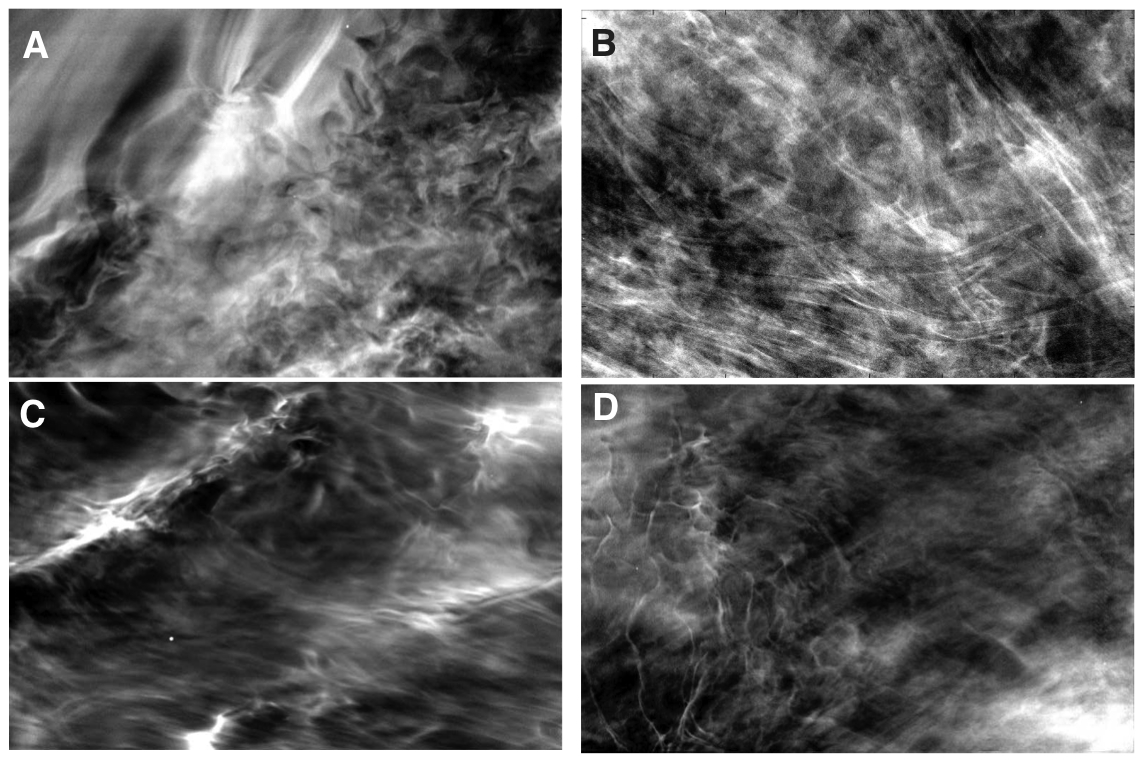}
\caption{Four EBEX images, 4.1 x 3.3 km FOV, illustrating additional intriguing dynamics of currently unknown origin including extremely fine-scale features as in Figure 4.}
\label{fig:other_dynamics_2}
\end{figure}

%
%


\end{document}